\documentclass[pdftex,singlecolumn,epjc3]{svjour3}          

\pdfoutput=1 

\usepackage{mathtools}
\usepackage{amsmath}
\usepackage{cancel}
\usepackage{dcolumn}
\usepackage{esvect}
\usepackage[x11names]{xcolor}
\usepackage{soul}

\RequirePackage{graphicx}
\RequirePackage{mathptmx}      
\RequirePackage{flushend}
\RequirePackage[numbers,sort&compress]{natbib}
\RequirePackage[colorlinks,citecolor=blue,urlcolor=blue,linkcolor=blue]{hyperref}

\newcommand{\CMO}{CaMoO$_4$}
\newcommand{\enCMO}{$^{48\textrm{depl}}$Ca$^{100}$MoO$_4$}
\newcommand{\zerodbd}{$0\nu\beta\beta$}
\newcommand{\twodbd}{$2\nu\beta\beta$}

\newcommand{\Mo}[1]{$^{#1}$Mo}
\newcommand{\Ge}[1]{$^{#1}$Ge}
\newcommand{\Xe}[1]{$^{#1}$Xe}
\newcommand{\Se}[1]{$^{#1}$Se}
\newcommand{\Te}[1]{$^{#1}$Te}

\newcommand{\Ca}[1]{$^{#1}$Ca}
\newcommand{\Bi}[1]{$^{#1}$Bi}
\newcommand{\Tl}[1]{$^{#1}$Tl}

\begin{document}

\newcolumntype{d}[1]{D{.}{\cdot}{#1} }
\title{First Results from the AMoRE-Pilot neutrinoless double beta decay experiment}


\author{V.~Alenkov\thanksref{FOMOS} 
        \and
        H.~W.~Bae\thanksref{KNU} 
        \and
        J.~Beyer\thanksref{PTB} 
        \and
        R.~S.~Boiko\thanksref{INR} 
        \and
        K.~Boonin\thanksref{Nakhon} 
        \and
        O.~Buzanov\thanksref{FOMOS} 
        \and
        N.~Chanthima\thanksref{Nakhon} 
        \and
        M.~K.~Cheoun\thanksref{Soongsil} 
        \and
        D.~M.~Chernyak\thanksref{INR} 
        \and
        J.~S.~Choe\thanksref{IBS} 
% ...
}
%
\thankstext{e2}{e-mail: tirstrike@snu.ac.kr}
\thankstext{e1}{e-mail: yhk@ibs.re.kr}

\institute{Center for Underground Physics, Institute of Basic Science (IBS), Daejeon 34126,
Republic of Korea\label{IBS}
% ...
\and
          JSC FOMOS-Materials, Moscow 107023, Russia\label{FOMOS}
          \and
          Department of Physics, Kyungpook National University, Daegu 41566, Republic of Korea\label{KNU}
          \and
          Physikalisch-Technische Bundesanstalt (PTB), Abbestr. 2-12, Berlin, 10587, Germany
\label{PTB}
          \and
          Institute for Nuclear Research, 03028 Kyiv, Ukraine\label{INR}
          \and
          Nakhon Pathom Rajabhat University, Nakhon Pathom 73000, Thailand\label{Nakhon}
          \and
         Department of Physics, Soongsil University, Seoul 06978, Republic of Korea\label{Soongsil}
% ...
}

\date{Received: date / Accepted: date}

\maketitle

\begin{abstract}
The Advanced Molybdenum-based Rare process Experiment (AMoRE) aims to search for neutrinoless double beta decay (0$\nu\beta\beta$) of \Mo{100} with $\sim$100 kg of \Mo{100}-enriched molybdenum embedded in cryogenic detectors with a dual heat and light readout. At the current, pilot stage of the AMoRE project we employ six calcium molybdate crystals with a total mass of 1.9~kg, produced from \Ca{48}-depleted calcium and \Mo{100}-enriched  molybdenum (\enCMO{}). The simultaneous detection of heat~(phonon) and scintillation~(photon) signals is realized with high resolution metallic magnetic calorimeter sensors that operate at milli-Kelvin temperatures. This stage of the project is carried out in the Yangyang underground laboratory at a depth of 700~m. We report first results from the AMoRE-Pilot \zerodbd{} search with a 111~kg$\cdot$d live exposure of \enCMO{} crystals. No evidence for \zerodbd{} decay of \Mo{100} is found, and a upper limit is set for the half-life of 0$\nu\beta\beta$ of \Mo{100} of $T^{0\nu}_{1/2} > 9.5\times10^{22}$~y at 90\% C.L.. This limit corresponds to an effective Majorana neutrino mass limit in the range $\langle m_{\beta\beta}\rangle\le(1.2-2.1)$ eV.

\end{abstract}

\section{Introduction}

Neutrinos, which correspond to a major portion of the elementary particles comprising the Universe, are still poorly understood.  
In the standard model (SM) of particle physics,  
neutrinos are assumed to be massless particles with $1/2$ spin. However, the recent discovery of neutrino oscillations~\cite{Fukuda1998atmosphericOscillation,Cleveland1998Homestake, Ahmed2001solarOscillation,fogli2012} demonstrates that the flavour states of neutrinos are linear combinations of mass eigenstates characterized by the Pontecorvo-Maki-Nakagawa-Sakata~(PMNS) matrix~\cite{pontecorvo1958sov,Maki1962PMNS}. These results clearly demonstrate that neutrinos have non-zero mass; the mass eigenstates' different time evolutions produce the observed oscillations between different flavour eigenstates.

The differences of the squares of the mass eigenvalues are
experimentally determined from the oscillation frequencies of solar, atmospheric, accelerator and reactor neutrinos~\cite{Fukuda1998atmosphericOscillation, Cleveland1998Homestake,Ahmed2001solarOscillation, tanabashi2018}. However, the absolute mass scale still remains unknown. Moreover, it has still not been determined if  neutrinos are Majorana particles, in which case neutrinos are their own antiparticles, or Dirac particles, in which case the neutrinos and antineurinos are distinct. This is a fundamental feature of the SM that remains unanswered~\cite{vergados2016review}.

In \textit{two-neutrino} double beta decay (\twodbd{}), expressed as
\begin{equation}
(Z,A)\rightarrow(Z+2,A)+2 \beta^- + 2 \bar{\nu}_e~,
\label{equation_0nbb}  
\end{equation}

\noindent
there is simultaneous emission of two electrons and two anti-neutrinos.  These processes have been observed for a number of nuclides~\cite{saakyan20132nbb,BARABASH201552}. 
On the other hand, \textit{neutrinoless} double beta decay~(\zerodbd{}), which is expressed as 

\begin{equation}
(Z,A)\rightarrow(Z+2,A)+2 \beta^- ~,
\label{equation_0nbb}  
\end{equation}

\noindent
is a hypothetical process that is allowed if neutrinos are Majorana particles with a non-zero mass. Since no anti-neutrinos are emitted in this process, it violates lepton number conservation by two units~\cite{vergados2016review,rodejohann_dbd,Vergados_2012, gomez2012search} and the total available decay energy is shared by the two electrons, with a negligible amount transferred to the daughter nucleus. Consequently, \zerodbd{} decay would produce a distinct narrow peak at the end-point of the broad electron energy spectrum of the \twodbd{} decay. 

The SM is invariant under a U(1) gauge transformation, leading to the lepton number conservation~\cite{weinberg1995quantum}. Searching for \zerodbd{} decay is currently the only plausible experimental technique to test the Majorana nature of neutrinos and the validity of lepton number conservation. The discovery of this decay would provide major new insights into the nature of beyond-the-SM physics. 

In general, the \zerodbd{} decay rate can be expressed as 

\begin{equation}
    [T^{0\nu}_{1/2}]^{-1} = G_{0\nu} \left| M_{0\nu} \right|^2 \left|\eta\right|^2~,
\label{equation_Thalf0nbb}
\end{equation}

\noindent
where $T^{0\nu}_{1/2}$ is the half-life of the \zerodbd{} decay, $G_{0\nu}$ is the phase space factor for the decay~\cite{Kotila2012PhaseSpace,mirea2015values}, $M_{0\nu}$ is the nuclear matrix element~\cite{Engel2017Status}, and $\eta$ is a parameter that depends on the mechanism of the process that mediates the lepton number violation process. If the process is mediated by light neutrino exchange, $\eta = \langle m_{\beta\beta}\rangle$, which is the effective Majorana mass expressed as the coherent sum 

\begin{equation}
\langle m_{\beta\beta}\rangle = \left| \sum^3_{i=1} U^2_{ei} m_i \right|~,
\end{equation}

\noindent
where $U_{ei}$ are PMNS matrix elements, and $m_i$ are the neutrino mass eigenstates. Note that Eq.~(\ref{equation_Thalf0nbb}) is the standard interpretation of the general lepton number violation processes. Other mechanisms that lead to \zerodbd{} at the same order strength as light neutrino exchange have been proposed~\cite{deppisch2012neutrinoless}. Therefore, a full understanding of the \zerodbd{} process would require \zerodbd{} measurements for more than one nuclide to distinguish between the different possibilities. In any case, a \zerodbd{} discovery would be a clear demonstration of the Majorana nature of neutrinos and of the lepton number violation. Moreover, the nuclear matrix values determined using different theoretical calculations differ by factors as large as two~\cite{Engel2017Status}. These uncertainties also call for \zerodbd{} search experiments that use different nuclides~\cite{Giuliani2018multi}.

To date, $0\nu\beta\beta$ decay has not been observed; the most sensitive experiments give lower limits on decay half lives at the level of $10^{24}-10^{26}$ y. The current best limit on any \zerodbd{} decay rate was established for \Xe{136} by the KamLAND-Zen collaboration, which used xenon-loaded liquid scintillator in a balloon and set a half-life limit of \zerodbd{} at $1.07\times~10^{26}$~y~\cite{kamland2016mass}. Germanium detector arrays made of \Ge{76} are used by GERDA~\cite{agostini2017background} and the \textsc{Majorana-Demonstrator}~\cite{abgrall2014majorana,alvis2019search}. A liquid xenon time projection chamber~(TPC) is employed by EXO-200~\cite{exo200_2018search}. CUORE searches for \Te{130} \zerodbd{} decay using thermal calorimeters operating at mK temperatures~\cite{alfonso2015search}. Recently, CUPID-0 carried out a \Se{82} \zerodbd{} decay search using the simultaneous heat and scintillation detection technique~\cite{cupid2018first}. 
NEMO used electron trackers and plastic scintillators for \zerodbd{} decay searches with various isotopes. 
The most recent version of the experiment, NEMO-3, obtained the most stringent half-life limit on \Mo{100} \zerodbd{} decay of $1.1\times 10^{24}$~y~\cite{arnold2015results}. Some of the most sensitive \zerodbd{} decay search experiments are summarized in Table~\ref{table_experiments}.

The Advanced Molybdenum-based Rare process Experiment (AMoRE)~\cite{amore_tdr,bhang_AMoRE} aims to search for \zerodbd{} decay of \Mo{100} using  a simultaneous heat and scintillation detection technique with crystals operating at mK temperatures. In the present stage of the project, AMoRE-Pilot, calcium molybdate~(\CMO{}) crystals are used as the active target elements in a detector concept where the source material is also the detection medium~\cite{Annenkov_CMO_property,cskang2017sst,jo2018jltp}. 
The crystals are \Mo{100}-enriched and \Ca{48}-depleted~(\enCMO{}). Since \Ca{48} is a double beta decaying isotope with a $Q$ value of 4271~keV, its \twodbd{} decay energy spectrum overlaps the 3034~keV energy region of interest~(ROI) for AMoRE's \zerodbd{} decay search, producing an irreducible background at the level of $\sim$0.01 counts/(keV$\cdot$kg$\cdot$y)~\cite{Annenkov_CMO_property}. Thus, the amount of \Ca{48} must be reduced.

\begin{table*} [t] 
\centering
\caption{A summary of the most sensitive \zerodbd{} decay search experiments. The target isotopes and the current half-life limits are shown}
\label{table_experiments}
\begin{tabular*}{\textwidth}{@{\extracolsep{\fill}}lccc@{}}
\hline
Experiment & Target isotope & Exposure (kg$\cdot$y) & $T_{1/2}$~limit (y) \\
\hline
KamLAND-Zen~\cite{kamland2016mass}  &  \Xe{136}  & 504  & 1.07$\times 10^{26}$  \\
GERDA~\cite{agostini2017background} &  \Ge{76}   & 34.4 & 5.3$\times 10^{25}$  \\
\textsc{Majorana}~\cite{alvis2019search}    &  \Ge{76}  & 26.0  & 2.7$\times 10^{25}$ \\
EXO-200~\cite{exo200_2018search}    &  \Xe{136}  & 178  & 1.8$\times 10^{25}$ \\
CUORE~\cite{alfonso2015search}      &  \Te{130}  & 9.8  & 4.0$\times 10^{24}$ \\
CUPID-0~\cite{cupid2018first}       &  \Se{82}   & 1.83 & 2.4$\times 10^{24}$   \\
NEMO-3~\cite{arnold2015results}     &  \Mo{100}  & 34.3 & 1.1$\times 10^{24}$   \\
\hline
\end{tabular*}
\end{table*}

Six \enCMO{} crystals with a total mass of 1.9~kg are currently in operation at the 700-meter-deep Yangyang underground laboratory (Y2L) located in South Korea. A cryogen-free dilution refrigerator~(CF-DR) is used to cool down the AMoRE-Pilot detector array to 10-20~mK, including sensors for heat and scintillation detection. The data sample used for the results reported here corresponds to a 52.1~kg$\cdot$d live exposure of $^{100}$Mo.  

\section{Experimental Setup and Data Acquisition}

\begin{figure} 
\centering
\includegraphics[width=8.4cm]{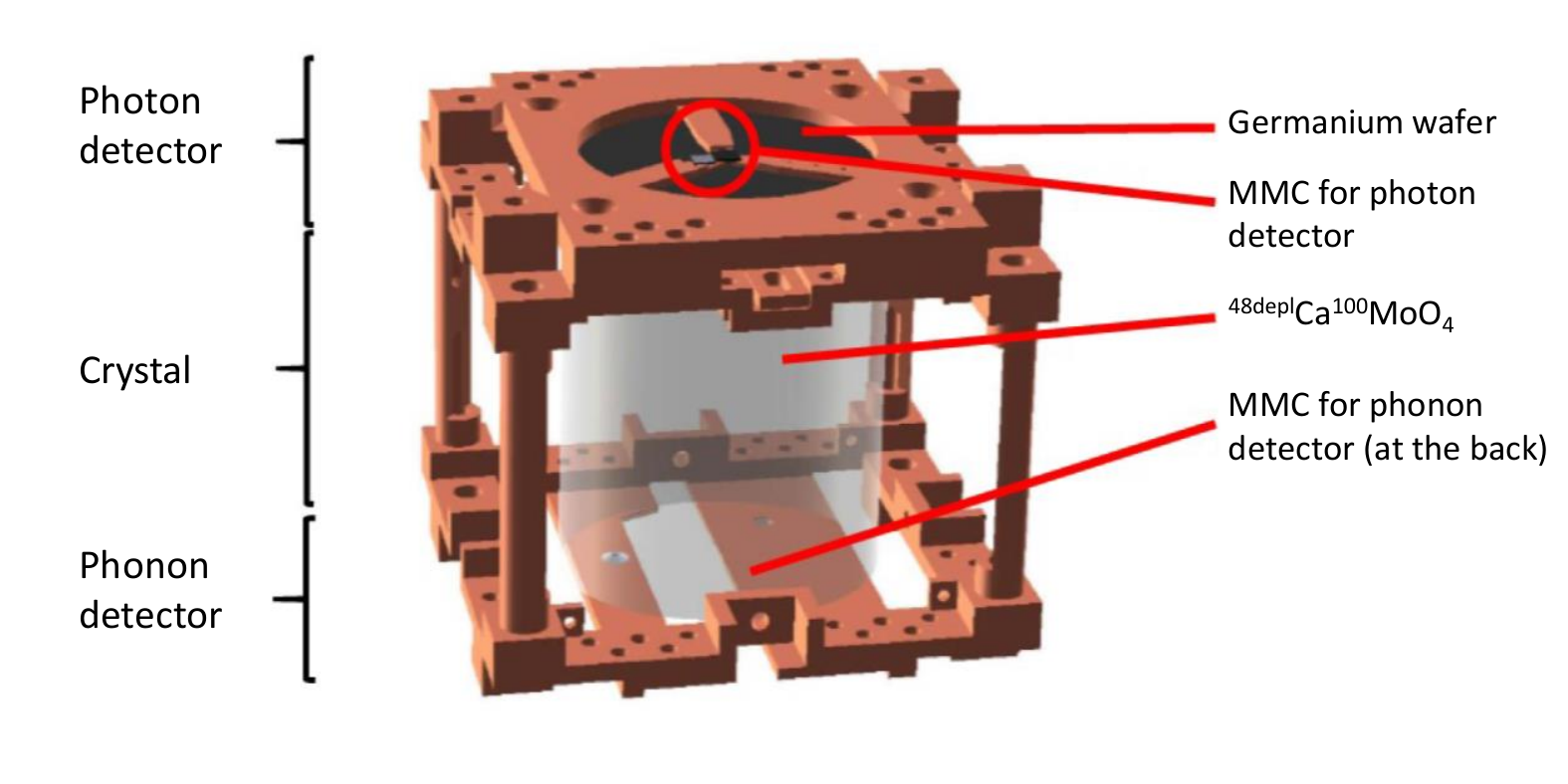}
\caption{A detector module for AMoRE-Pilot experiment. It is comprised of three main parts: a doubly enriched \enCMO{} crystal, a photon detector, and phonon detector. Figure from ~\cite{ikim2017sst}.} 
\label{figure_detectorModule}
\end{figure}

The AMoRE-Pilot detector modules are comprised of three main components, as shown in Figure~\ref{figure_detectorModule}: a \enCMO{} scintillating crystal, a phonon sensor based on a metallic magnetic calorimeter (MMC) that measures the temperature rise of the crystal induced by radiation absorption, and a photon detector with an MMC sensor that detects the amount of scintillation light produced in the crystal~\cite{gbkim_trans,KIM2017105,ikim2017sst}.

\begin{figure} 
\centering
\includegraphics[width=8.4cm]{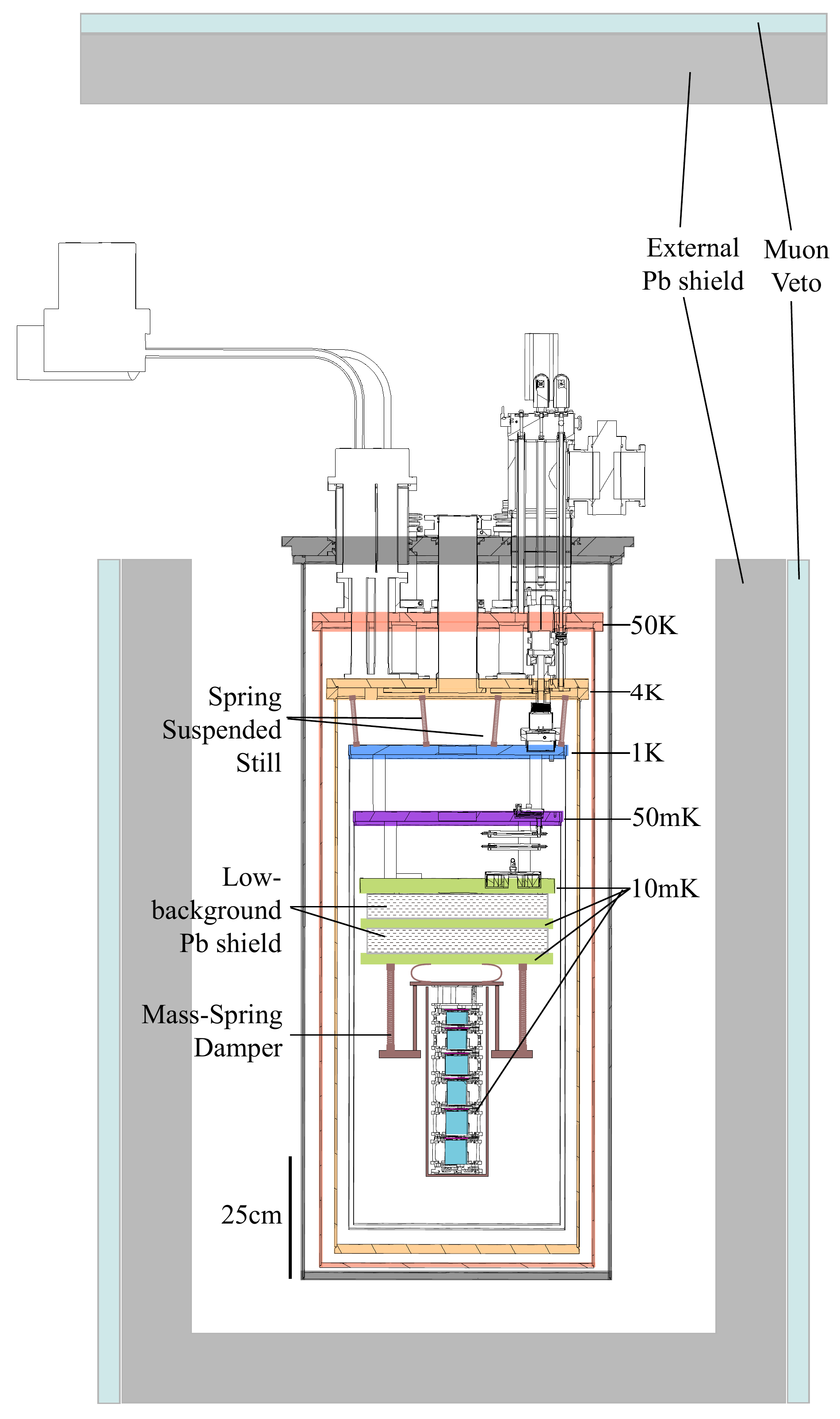}
\caption{Schematic diagram of the AMoRE-Pilot detector system.}
\label{figure_mano}
\end{figure}

The crystals are assembled inside a high radiopurity, high conductivity copper frame that also functions as a thermal bath. Three conically shaped polyether-ether-ketone~(PEEK) spacers placed beneath the crystals provide support with minimal thermal losses. Four copper tabs on the top of each frame firmly press down the crystals and keep them stationary. Polytetrafluoroethylene~(PTFE) spacers are installed between the holders and the crystal to reduce heat losses. The crystals are elliptical cylinders of different sizes, with masses that range from 196~g to 390~g.

The phonon sensor is situated at the bottom of the detector-module copper frame. A 400~nm thick gold film deposited on the lower surface of the crystal serves as a phonon collector. This is connected to an MMC by 25 annealed gold wires~\cite{yhkim2004}. The MMC exploits the paramagnetic nature of erbium ions in a gold host (Au:Er)~\cite{enss_MMC}. The devices were micro-fabricated on a Si wafer by sputtering a planar niobium coil with a layer of Au:Er sensor material immediatly above the coil~\cite{fleischmann2009metallic, wsyoon2015}. A persistent current in the superconducting coil generates a magnetic field in the Au:Er sensor material, and energy deposited in the crystal produces a temperature increase in the Au:Er that alters its magnetization.  This results in a change in the coil current that is measured by a superconducting quantum interference device (SQUID)~\cite{cskang2017sst}. Because of the high sensitivity of the MMC sensors, the detector setup is able to measure precisely the temperature increase resulting from any type of radiation absorption, and particle detection with very high energy resolution is achievable~\cite{ikim2017sst,Kempf2018}.

A detachable photon detector~\cite{hjlee2015} is installed at the top of the copper frame. The photon detector is comprised of a 2-inch diameter, 300~$\mu$m thick germanium wafer that serves as a scintillation light absorber; an MMC sensor measures the phonon energy generated inside the germanium wafer. 
With this arrangement, we measure the phonon signals and scintillation light signals simultaneously; by comparing the relative amplitudes of the simultaneous signals,  we can discriminate between the unwanted non-$\beta/\gamma$-generated events and the $\beta/\gamma$-like  events of interest~\cite{gbkim_trans,KIM2017105,ikim2017sst}.

The AMoRE-Pilot experiment involved five data-taking runs between 2015 and 2017~\cite{jo2018jltp}. The goals of the first four runs were to: realize the measurement setup with a refrigerator system and a lead shield system in the underground laboratory; investigate the noise and the background levels; and improve the detector performance. During these research and development stages, we used a CF-DR with a 7~mK minimum temperature in a newly built laboratory room in Y2L. As shown in Figure~\ref{figure_mano}, we also assembled a detector array comprised of several detector modules below the 10~mK plate. For the first five runs, five \enCMO{} crystal detector modules with a total mass of 1.547~kg (0.727~kg of $^{100}$Mo) were operated. During this R\&D phase, we significantly improved the detectors performance. We installed a 15~cm-thick lead shield to cover the top and the sides of the CF-DR to suppress background radiation from external sources. An active muon veto system~(MVS) was used to tag and reject muon-induced background events. Vibration noise from the pulse tube refrigerator of the CF-DR, initially the dominant source of signal noise, was reduced  by a few orders of magnitude by means of a two-stage vibration isolation system consisting of a spring suspended still~(SSS) with eddy current dampers~(ECDs)~\cite{Lee2018} and a mass-spring damper that mechanically isolates the detector tower from the 10~mK CF-DR mixing chamber plate~(See Figure~\ref{figure_mano}). During the fifth run, the SSS mechanism was not fully operational because of damage to one of its magnets. However, it still had outstanding vibration reduction performance that improved the noise conditions by several orders of magnitude~\cite{Lee2018}. The SSS system was in full operation for the subsequent runs.

The fifth run of AMoRE-Pilot, the main data collection run of the experiment, started operation in August 2017. For this, we added a sixth \enCMO{} crystal of mass 340~g, thereby increasing the total crystal mass to 1.887~kg (0.886~kg of $^{100}$Mo). The six detector modules were arranged in a single tower inside a superconducting lead shield. The detector tower and shield were maintained at a temperature of 20~mK throughout the data collecting period. 

Data were acquired by a 12 channel 18-bit ADC, including six phonon and six photon channels. A continuous stream of data was saved and later triggered off-line via the application of a Butterworth filter~\cite{Selesnick1998butterworth,Kim2018}. The full dynamic range of the measurement was $\pm$ 3~V, and limited by a differential amplifier. The sampling rate of the DAQ was set to 100~kHz, which is sufficient for pulse shape discrimination~(PSD) of different types of background events based on the rise time of the phonon signals. During this phase of the experiment, we collected a 69.0~kg$\cdot$d $^{100}$Mo exposure. The six detector channels and the cryogenic system operated reliably throughout the approximately 1000 hour measurement period.

\section{Primary Data Processing} \label{ch:primary}

A Butterworth bandpass filter was applied to both of the original phonon and photon signals to enhance the signal-dominant frequency band and suppress noise-dominant frequency bands~\cite{Kim2018}. We evaluated many sets of passbands and chose the set that provided the highest signal-to-noise ratio~(SNR) for each channel. The filtered signals were also well suited for effective triggers~\cite{Kim2018}. The energy threshold was between 2 and 5~keV depending on the crystal; the off-line trigger levels were set to several tens of keV for the \zerodbd{} decay search analysis; the $Q$-value of  \Mo{100} \zerodbd{} decay is $\sim$3~MeV.

To determine the phonon and photon signal amplitudes for each event, the filtered signals were fitted by a least squares method that minimized the sum of squared residuals with a scaled filtered template developed with 2615~keV gamma ray signals from \Tl{208} contaminants in the crystals. This is the highest naturally occurring intensive $\gamma$ peak and the closest one to the ROI.
For a signal vector $S_i$ in the time domain and the template signal vector $T_i$ of length $N$, scaled with the scale factor $A$, the squared error to the template~($SE$) is written as:
\begin{equation}
    SE = \Sigma_{i = 1}^N (S_i - A T_i)^2~.
\label{equation_squareError}
\end{equation}
The scale factor $A$ that minimizes $SE$ is the amplitude of the signal normalized to the template:
\begin{equation}
    A = \frac{\Sigma_{i = 1}^NS_i \cdot T_i}{\Sigma_{i = 1}^NT_i^2} ~.
\label{equation_LSQ}
\end{equation}
Here we assume that all the $\beta/\gamma$-signals have the same shape as the template, but with different amplitudes. Divergence from this assumption produces a small non-linearity in the amplitude-to-energy calibration. The amplitudes of non-$\beta/\gamma$-like events are also calculated with this method, but their energy scale is different because of the mismatch between their pulse shape and that of the template. 

Signal amplitudes can be influenced by a long term instability of the temperature. In the present system, it appears as a long-term thermal gain drift that can be corrected with frequent reference to events with a known energy. For AMoRE-Pilot, one of the six crystals, Crystal~2~(See Table~\ref{table_DP}), has a high internal $^{210}$Po $\alpha$ rate. These events were used as standard signals to provide gain stabilization signals for all six crystals. The application of the gain stabilization for one detector using events from another one was demonstrated in previous AMoRE-Pilot runs. For future runs, gain corrections will be improved with stabilization heaters installed on each crystal~\cite{arnaboldi2005stabilization}.

The energy calibration of each crystal was done with a $^{232}$Th $\gamma$ source rod placed in between the lead shield and the outer vacuum chamber of the cryostat for one day each week. The background spectrum was compared to the calibration spectrum for peak identification. We then made use of between 7 and 11 strong $\gamma$-peaks to determine the signal amplitude {\it vs} the energy calibration. 

\begin{figure} 
\centering
\includegraphics[width=8.4cm]{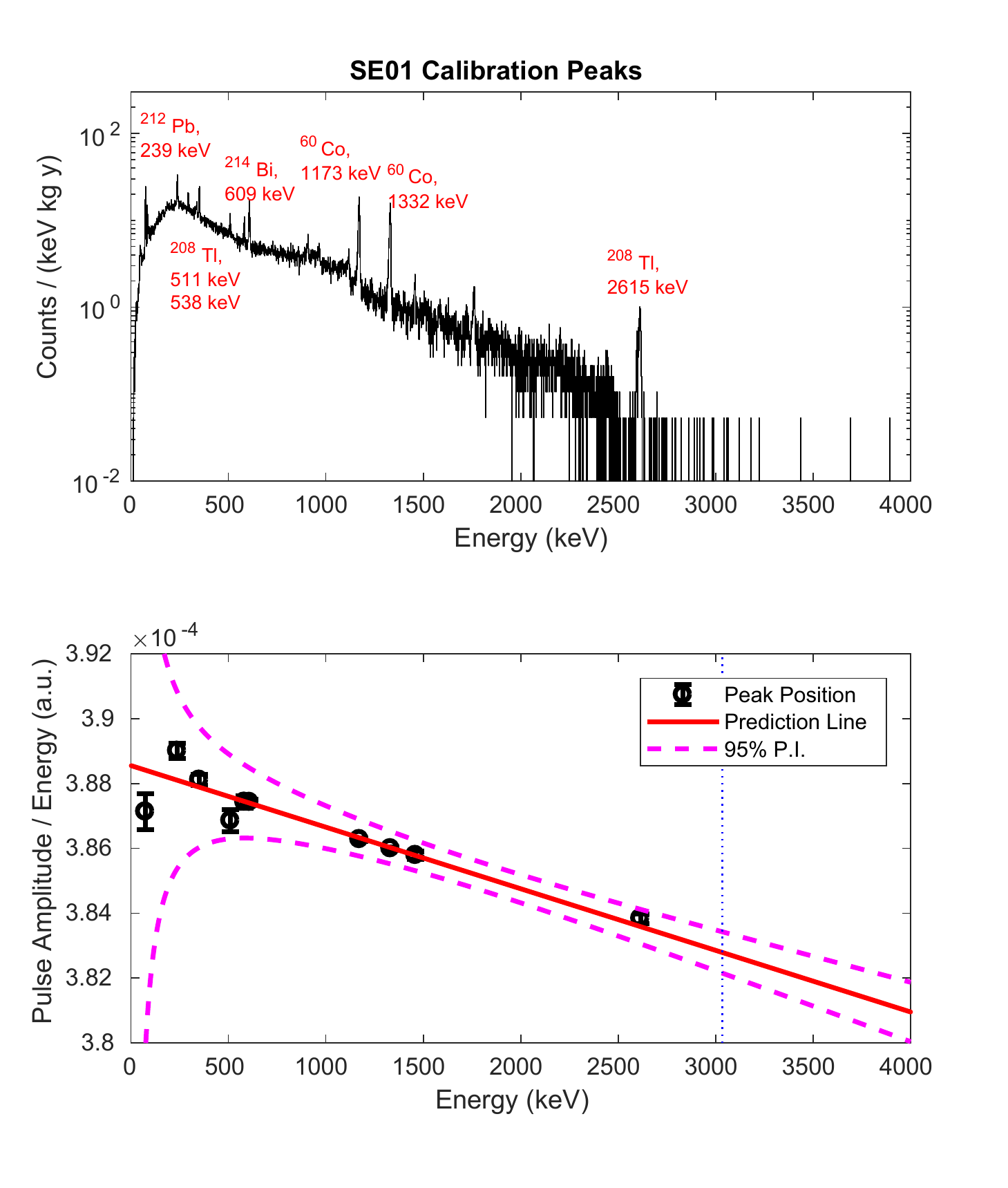}
\caption{\textbf{Top}: Energy spectrum of one \enCMO{} detector for a 60~day exposure  after the application of the event selection criteria,  without a calibration source. The most prominent $\gamma$ peaks are labelled. \textbf{Bottom}: A second-order polynomial calibration curve~(solid) with its 95\% P.I.~(dashed). The vertical dashed line in the bottom figure indicates the $Q$ value of \Mo{100} double beta decay. See Section~\ref{ch:selection} for a detailed explanation of the event selection methods. (Colour figure online)}
\label{figure_calibration}
\end{figure}

Our earlier work suggested that the amplitudes of $\beta/\gamma$ peaks follow a simple quadratic function of energy up to 2615~keV~\cite{KIM2017105,ikim2017sst}; we extrapolate this to higher energies. The suitability of a quadratic polynomial with no constant term for energies above 2615~keV was verified by fitting $\alpha$-peaks, including the 2311~keV $^{147}$Sm peak, with another quadratic polynomial function. Although $\alpha$- and $\beta/\gamma$-events generate pulses of different amplitudes due to the different signal shapes, good agreement between $\alpha$- calibration lines and a quadratic polynomial provides evidence that a similar detector response is adequate for $\beta/\gamma$-events with energies above 2615~keV.

\begin{figure} 
\centering
\includegraphics[width=8.4cm]{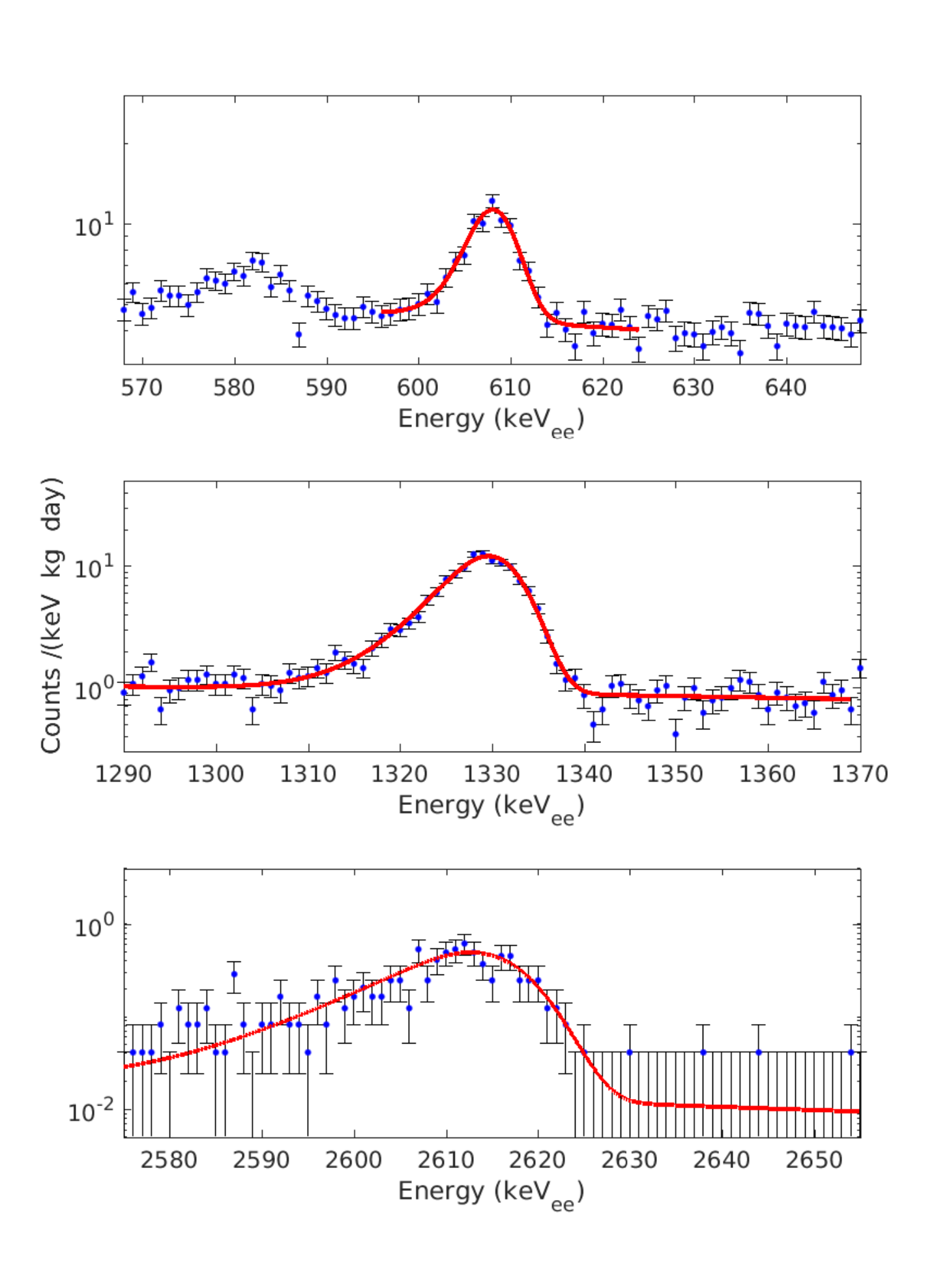}
\caption{$\gamma$ peaks from a background spectrum of one of the \enCMO{} detectors. \textbf{Top:} 609~keV from $^{214}$Bi. \textbf{Middle:} 1332~keV from $^{60}$Co. \textbf{Bottom:} 2615~keV from $^{208}$Tl. Solid lines show the best fit of the gamma peaks by exponentially modified Gaussian (see text) with an exponential function to describe the continuous background. The skewness increases with the energy. (Colour figure online)}
\label{figure_EMG_1332keV}
\end{figure}

Figure~\ref{figure_calibration} shows a background energy spectrum from one of the crystals that shows signal amplitudes that cover the energy range below 2615~keV, and displays a non-linear behaviour and deviations from the calibration. The solid line shows the calibration. The negative slope reflects the non-linear term. The peak positions are in good agreement with the 95\% calibration prediction interval~(P.I.) shown as dashed lines,  indicating that the energy of a signal can be estimated using the calibration and error over a wide energy region. Uncertainties in the energy calibration are included in the likelihood marginalization~\cite{geisser2017predictive}.

Figure~\ref{figure_EMG_1332keV} shows shapes of $\gamma$ peaks with energies 609, 1332 and 2615~keV in the data accumulated with one of the \enCMO{} detectors. The 1332~keV peak originated from $^{60}$Co in the outer vacuum chamber. The solid curves are the fits of the energy~($E$)-dependent model function of the detector response to mono-energetic $\gamma$-rays in the presence of exponential backgrounds:
\begin{equation}
\begin{aligned}
	f(E;A,\mu,&\sigma,\lambda,a,b)= \\
		& \xi \cdot f_\textrm{EMG}(E;\mu,\sigma,-\lambda) + a \cdot \textrm{exp}(-b \cdot E)~.
\end{aligned} 
\label{equation_EMGExp}
\end{equation}
Here $f_\textrm{EMG}(E;\mu,\sigma,-\lambda)$ is an exponentially modified Gaussian with a mean $\mu$, standard deviation $\sigma$ and negative exponential component $-\lambda$:

\begin{equation} \label{equation_EMG_2}
f_\textrm{EMG}(E;\mu,\sigma,-\lambda) = -\frac{\lambda}{\sqrt{\pi}} \textrm{exp}\Bigg(-\frac{\lambda}{2}\Big(2 \mu - \lambda \sigma^2 - 2 E\Big)\Bigg) \int^\infty_E \textrm{exp}(-t^2) \textrm{d}t~,
\end{equation}

\noindent
where $\xi$ is the signal strength parameter, and $a$ and $b$ are parameters from an exponential representation of the continuous background. The EMG function is used to model the low energy tails of $\gamma$ and $\alpha$ peaks caused by incomplete phonon collection in the detectors. The skewness $1/ \lambda$ was set to be proportional to the mean ($\mu$) of the model function. Fits using this prescription to many peaks with a single value for $E/\lambda$ give good fits to the measured peaks for all the crystals and is in the model function for \zerodbd{} events.

\section{Event Selection} \label{ch:selection}

\begin{figure} 
\centering
\includegraphics[width=8.4cm]{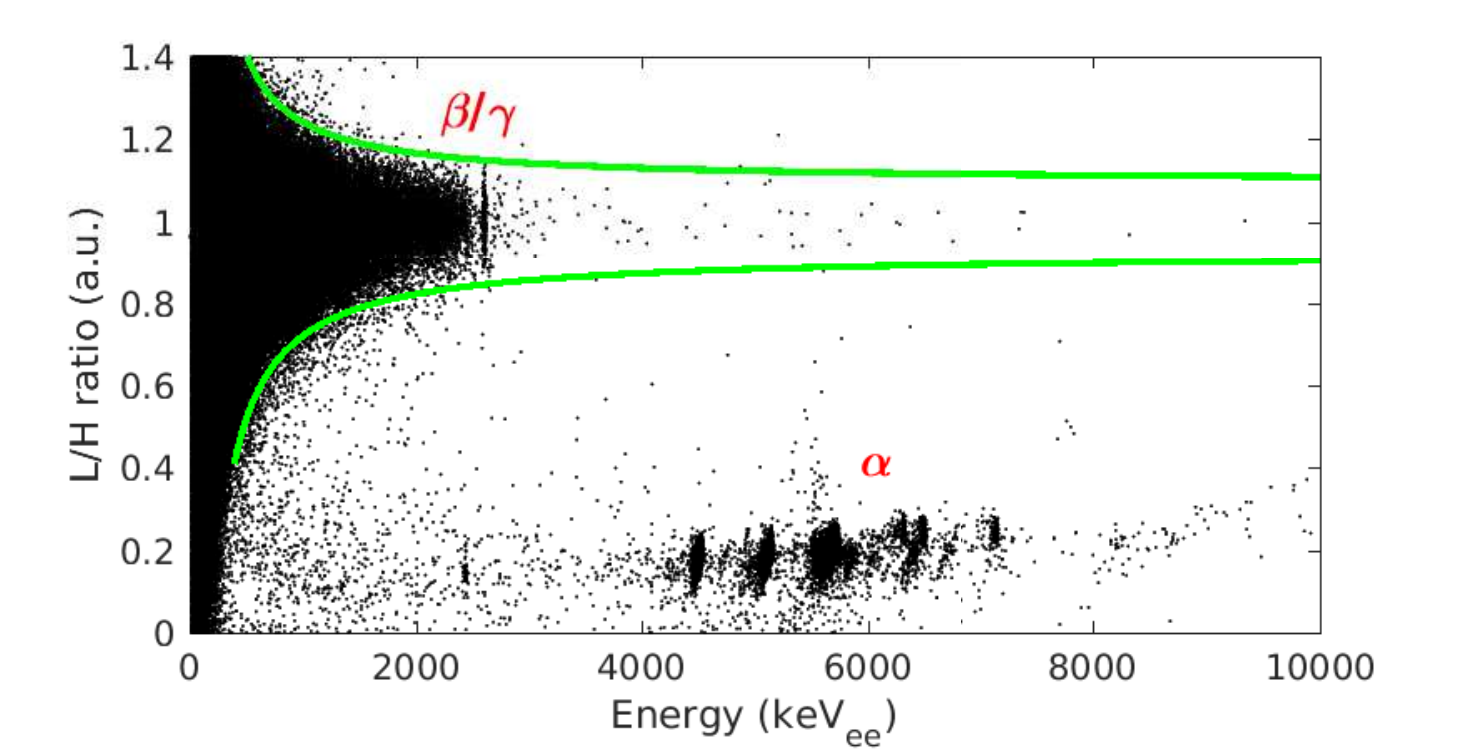}
\caption{Scatter plot of the light yield $vs$ energy deposited in the phonon channel for one crystal. The $\beta/\gamma$ band and the $\alpha$ clusters are clearly separated, as labelled in the figure.}
\label{figure_LH}
\end{figure}

We used a variety of event selection methods to improve the signal to background ratio in the ROI. The selection criteria are based on operational status, event-by-event identification, and coincidence tagging.

The operational status selection excludes the data-taking periods where  measurements were inactive or unstable. The rejected periods include times for system maintenance, calibration, and periods of excessive noise and sensor instabilities; and dominate the total dead-time, resulting in a live-time that is 84\% of the total data-taking time interval.

\begin{figure} 
\centering
\includegraphics[width=8.4cm]{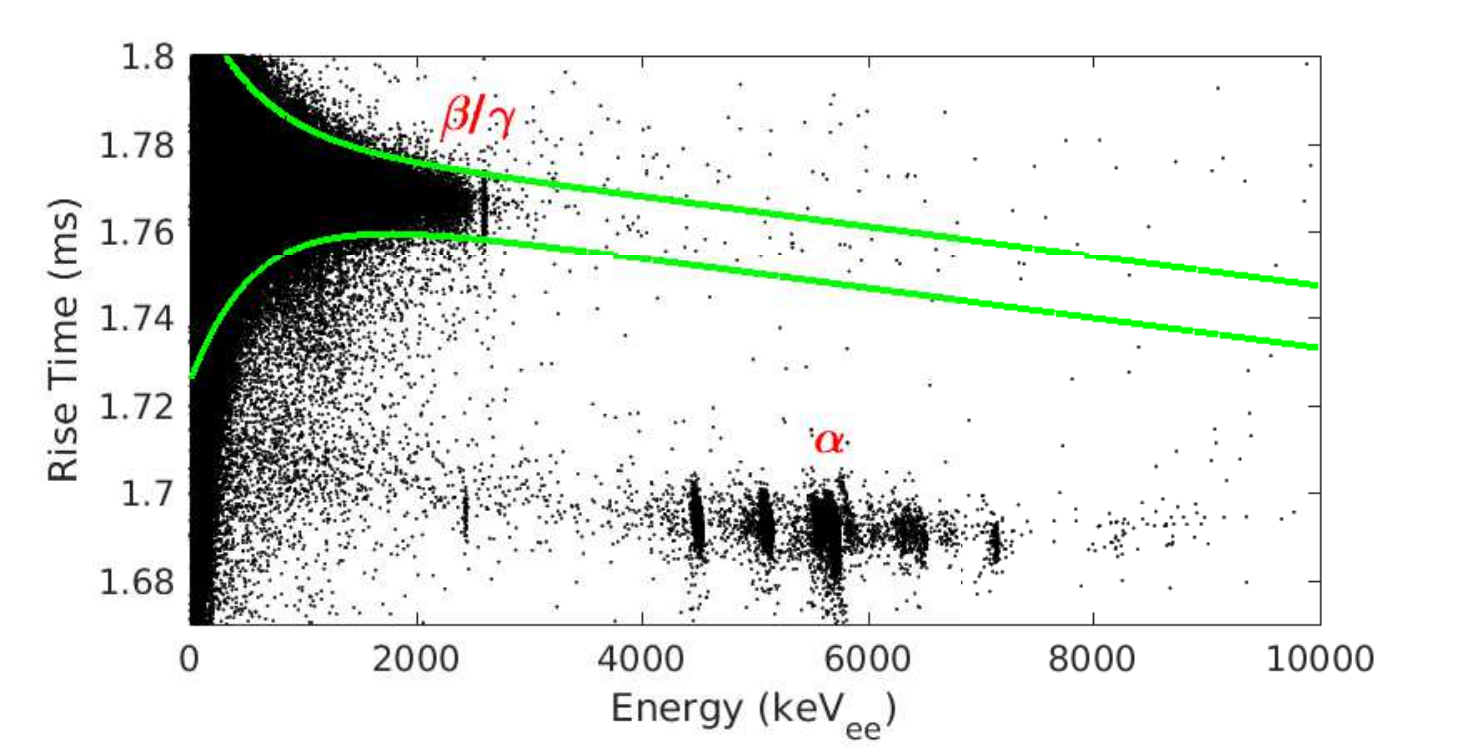}
\caption{Scatter plot of the rise time $vs$ energy deposited in a phonon channel. The $\beta/\gamma$ band and the $\alpha$ clusters are clearly separated, as labelled in the figure.}
\label{figure_RT}
\end{figure}

The event-by-event identification selection criteria discriminate $\beta/\gamma$-induced events from other background signals: $\alpha$-induced signals; SQUID resets, which appear as a sudden change in DC value by integer multiples of a unit magnetic flux quantum; anomalous signals that have poor template fits as indicated by  large minimum $SE$ values; and high energy events that exceed the dynamic range.
These were implemented by using two particle identification~(PID) techniques. One is from the relative amplitude ratio between light and heat signals (L/H ratio) in the simultaneous phonon-scintillation measurement and the other is the pulse shape discrimination~(PSD) of the heat signals.

The underlying concept of the L/H ratio method is that the amount of scintillation light emitted by a scintillator as a result of the interaction with a charged particle depends inversely on its stopping power~\cite{TRETYAK201040}. 
An event induced by a $\beta/\gamma$ deposit in a crystal yields a larger light signal than one induced by an $\alpha$ with the same energy, resulting in different L/H ratios. Figure~\ref{figure_LH} displays a 2D scatter plot for all events after the operational status and non-physical signals rejections, where clear $\beta/\gamma$ and $\alpha$ separation is evident.

The pulse shape of the phonon signal can also be used for event identification. The \CMO{} bolometric signal rise time~(RT), the time difference between 10\%$-$90\% of the pulse maximum, is different for $\beta/\gamma$- and $\alpha$-induced events~\cite{arnaboldi2011psd,GIRONI2013psd,gbkim_trans,ikim2017sst}. In both the L/H~ratio and RT distributions, some events appear in between the two distinct groups. These events are mostly caused by random signal overlaps (pileups) and/or poorly reconstructed signal shapes.

Using either L/H or RT values for event identification,  a discrimination power (DP) for background rejection can be defined as: 
\begin{equation}
    \textrm{DP} = \frac{|\mu_{\beta/\gamma}-\mu_{\alpha}|}{\sqrt{\sigma_{\beta/\gamma}^2 +\sigma_{\alpha}^2}}~,
\label{equation_DP}
\end{equation}

\noindent
where the $\mu$'s and $\sigma$'s are the mean values and the standard deviations of those events distributions, respectively.  
For the distributions of the parameters $\mu$ and $\sigma$, the two techniques have different energy dependence. Assuming the light yield per keV energy deposit in a crystal is independent of the deposited energy, we can infer a constant value for the $\mu_{\beta/\gamma}$ parameter of the L/H ratio. The validity of this assumption was established in previous reports~\cite{KIM2017105,ikim2017sst}. The other values$-\sigma$ for the L/H discriminator, and $\mu$ and $\sigma$ for the RT discriminator $-$depend on the energy.
The DP values of the two event identification parameters, determined for energies at the $Q$ value of \Mo{100} decay, are listed in Table~\ref{table_DP}.

\begin{table} [t] 
\centering
\caption{DP values and the effective selection efficiency~(See Section~\ref{section_eff}) by using both the L/H and RT cuts at the $Q$ value of \Mo{100} decay (3034~keV)}
\label{table_DP}
\begin{tabular*}{\columnwidth}{@{\extracolsep{\fill}}lllcl@{}}
\hline
Crystal (mass) & \multicolumn{1}{l}{DP$_{\textrm{L/H}}$} & \multicolumn{1}{l}{DP$_{\textrm{RT}}$} &  Selection efficiency [\%]\\
\hline
Crystal 1  (196~g)    &  7.07 & 18.0  & 91.2\\
Crystal 2  (256~g)    & 15.1  &  6.22 & 91.5 \\
Crystal 3  (350~g)    & 14.1  &  4.12 & 91.0 \\
Crystal 4  (354~g)    & 11.3  & 12.5  & 91.3 \\
Crystal 5  (390~g)    & 10.2  &  9.64 & 91.4 \\
Crystal 6  (340~g)    &  8.30 & 17.2  & 91.0 \\
\hline
\end{tabular*}
\end{table}

For $\beta/\gamma$ events, we assume a constant mean and dispersion of the L/H ratio. The measured distribution was divided into narrow, 100~keV energy slices between 465~keV and 2665~keV, and Gaussian fits were used to determine 24 different values of $\sigma_{\beta/\gamma}$.
The energy dependent upper and lower bounds, $g_{\pm}$, on the L/H ratio are parametrized by 

\begin{equation}
    g_{\pm}(E) = \pm c_1~\textrm{exp}( - c_2~E) + c_3 ~,
\label{equation_LYbound}
\end{equation}

\noindent
with $c_i$ values determined from fits to the L/H ratio distribution. The function determined from the measured values of $\mu \pm 2 \sigma_{\beta/\gamma}$ is shown as a green band in Figure~\ref{figure_LH}. The events inside the band are considered as $\beta/\gamma$ events. 

A similar method was applied to the RT parameter, as shown in Figure~\ref{figure_RT}. Since the pulse shape for the phonon channel has some energy dependence, a linear term was added to the mean and the dispersion of the RT distribution. Using $\mu \pm2\sigma_{\beta/\gamma}$  of the RT distribution as upper and lower limits, the characteristic function $h_{\pm}$ of the dispersion is:
\begin{equation}
    h_{\pm}(E) = \pm d_1~\textrm{exp}( - d_2~E) + d_3 E + d_4~,
\label{equation_RTbound}
\end{equation}

\noindent
where $d_i$ are fitting parameters.

\begin{table*} [h] 
\caption{\enCMO{} and $^{100}$Mo exposure of the AMoRE-Pilot experiment after each event selection method}
\label{table_Exposure}
\begin{tabular*}{\textwidth}{@{\extracolsep{\fill}}lccl@{}}
\hline
\multicolumn{1}{c}{Selection method} & \multicolumn{1}{l}{\enCMO{} exposure (kg$\cdot$d)}& \multicolumn{1}{l}{$^{100}$Mo exposure (kg$\cdot$d)} \\
\hline
 \hspace{3mm}Total Exposure      &  147 & 69.0 \\
 \hspace{8mm}+Operational status &  123 & 58.0 \\
 \hspace{8mm}+\Bi{212} tag       &  121 & 57.0 \\
 \hspace{8mm}+Trigger dead time  &  111 & 52.3 \\
 \hspace{8mm}+$\mu$-veto         &  111 & 52.2 \\
 \hspace{8mm}+Multiple hits      &  111 & 52.1 \\
\hline
\end{tabular*}
\end{table*}

The third class of event selection requirements, coincidence tags, removes multiple-hit events and pairs of decays correlated in time. While many background events can deposit energy across multiple locations, double beta decay events are contained within short ranges of the emitted electrons, confining the energy deposition to a single crystal. Thus a multiple-hit selection was applied to reject events with signals in more than one crystal within a 2~ms coincidence window. The MVS surrounding the external lead shield provided the capability for rejecting signals induced by cosmogenic muons. Approximately 2000~muons/day are detected in the MVS.  A veto period of 100~ms was applied following each muon signal to reject any coincident signals in the crystals. During the pilot stage, the MVS did not provide the full 4-$\pi$ coverage over the detector modules. The muon rejection efficiency was determined to be between 87\% and 92\%, depending on the crystal location. 
Finally, we reject background events from internal \Tl{208} $\beta$ decays, which have a $Q$ value of 4999~keV, by tagging the prior
6207~keV emission from the associated $\alpha$ decay of \Bi{212}.
The time between the two decays is characterized by 3.05-minute half-life of \Tl{208} $\beta$ decay and, thus, a veto-time of 30~minutes was set following the \Bi{212} $\alpha$ detection. This dead time only applies to a single crystal. The good energy resolution and identification capabilities of  the crystals allowed the \Bi{212} events to be tagged with high specificity. With this method, the \Tl{208} rejection efficiency in the ROI was 97.5\%. In the present analysis, the dead time from the 30~minute \Bi{212} tagging was 1.7\%, as indicated in Table~\ref{table_Exposure}. Since the phonon signal duration is approximately 1~s, we applied a trigger dead time of 1~s for each of the triggered events. The exposures that survived each event selection requirement are summarized in Table~\ref{table_Exposure}.

\begin{figure} 
\centering
\includegraphics[width=8.4cm]{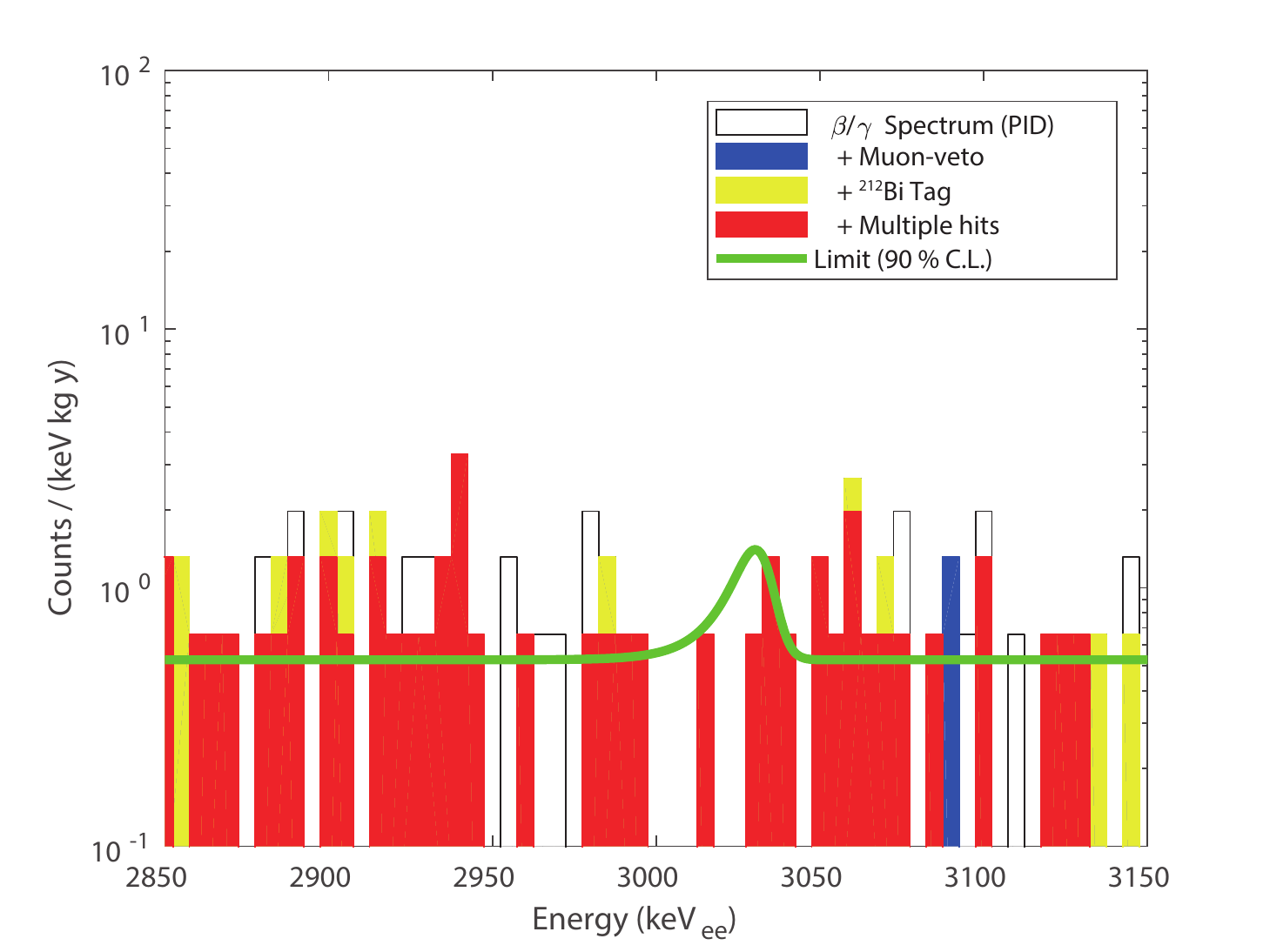}
\caption{The energy spectra gathered by \enCMO{} detectors with the 111~kg$\cdot$d exposure in the ROI before and after applying coincidence tagging techniques. Each spectrum represents the result after the cut criterion overlaying the previous accumulation. The solid green line is the fitted response function describing the flat background and the excluded \zerodbd{} peak corresponding to the 90\% confidence level~(C.L.) limit of $T_{1/2}^{0\nu} > 9.5\times10^{22}$ y. (Colour figure online)}
\label{figure_ROISpectrum}
\end{figure}

The energy spectrum in the $2850-3150$~keV ROI for the 111~kg$\cdot$d exposure that remains after the application of the event selection requirements and coincidence tagging vetoes is shown in Figure~\ref{figure_ROISpectrum}. The spectrum shape of the 50 events in the plot is approximated by a linear background function plus the EMG function for the \zerodbd{} peak. The width of this energy region was chosen to cover the total response function for  \zerodbd{} events while avoiding the influence of the 2615~keV \Tl{208} peak.

The systematic errors associated with the exposure include the uncertainties of mass measurement and $^{100}$Mo content estimations. The crystals were weighed with a precision of $\pm~0.5$~g. The  $^{100}$Mo content was determined to range from 95.0\% to 97.0\% ($\pm0.2$\%) by a series of ICP-MS measurements for the raw materials used for the crystals production. 
The systematic uncertainty for the $^{100}$Mo content corresponds to an almost negligible $\pm~0.057$~kg$\cdot$d.

\section{\zerodbd{} analysis}

\subsection{\zerodbd{} peak detection efficiency} \label{section_eff}

\begin{table}
\centering
\caption{Detection efficiency for the \zerodbd{} peak of \Mo{100} (\%)}
\label{table_Efficiency}
\begin{tabular*}{\columnwidth}{@{\extracolsep{\fill}}lll@{}}
\hline
\zerodbd{} peak absorption efficiency            & 81.6 $\pm~0.008$ \vspace{1mm}\\
Events selection efficiency                   & 91.2 $^{+0.1}_{-0.4}$ \\
\vspace{2mm} \\
Total \zerodbd{} peak detection efficiency   & 74.4 $^{+0.1}_{-0.3}$\vspace{1mm}\\
\hline
\end{tabular*}
\end{table}

Monte-Carlo simulations using the GEANT4 package~\citep{AGOSTINELLI2003250} for \zerodbd{} events generated with the DECAY0 code~\cite{Ponkratenko2000} were carried out to evaluate the efficiency for full energy deposition in the crystals. Taking the actual dimensions of the crystals into account for the simulations,  peak detection efficiencies of 79.3--82.4\% in the 3024--3044~keV energy region were found for 10$^5$~\zerodbd{} events generated in each crystal, even though the masses of the lightest and heaviest ones differ by nearly a factor of two. The largest uncertainty in the peak detection efficiency estimation is due to the crystal dimensions.  A simple scaling method is used to determine an uncertainty of 0.01\% in absolute detection efficiency. 

The efficiency for selecting \zerodbd{} events is further adjusted to account for the cuts used for $\beta/\gamma$ identification. We applied $\pm$2$\sigma$ cuts to select $\beta/\gamma$-induced events using both L/H and RT parameters. The selection efficiency for each cut is 95.45\%. After consideration of the correlation between the two parameters in the ROI, the selection efficiency was determined to be (91.2$^{+0.1}_{-0.4}$)\%.

The overall detection efficiency for the present experiment is ($74.4^{+0.1}_{-0.3}$)\% and summarized in Table~\ref{table_Efficiency}.

\subsection{Model of background in the ROI} \label{section_ROI}

\begin{table*} 
\centering
\caption{Parameters of the response function~(Eq.~(\ref{equation_EMG_2})) at the \zerodbd{} decay $Q$-value for each detector}
\label{table_Parameters}
\begin{tabular*}{\textwidth}{@{\extracolsep{\fill}}lllrc@{}}
\hline
Detector   & \multicolumn{1}{c}{$\mu_{0\nu\beta\beta}$} 
               & \multicolumn{1}{c}{$\sigma_{0\nu\beta\beta}$} 
                  & \multicolumn{1}{c}{$\tau_{0\nu\beta\beta}$}  
                     & \multicolumn{1}{c}{FWHM$_{0\nu\beta\beta}$}  \\
            & \multicolumn{1}{c}{[keV$_{\textrm{ee}}$]} 
               & \multicolumn{1}{c}{[keV$_{\textrm{ee}}$]} 
                  & \multicolumn{1}{c}{[keV$_{\textrm{ee}}$]}
                     & \multicolumn{1}{c}{[keV$_{\textrm{ee}}$]}    \\ 
\hline 
Crystal 1 (196~g)    & 3034 $\pm~1.1$ &  4.7 $\pm~0.26$ &  7.2  $\pm~0.0003$ 
             & 15.2           \vspace{1mm} \\
Crystal 2 (256~g)   & 3034 $\pm~0.5$ &  3.2 $\pm~0.30$ &  5.2  $\pm~0.0003$ 
             & 10.6           \vspace{1mm} \\
Crystal 3 (350~g)   & 3034 $\pm~1.6$ &  4.5 $\pm~0.22$ & 10.7   $\pm~0.0001$  
             & 17.3           \vspace{1mm} \\
Crystal 4 (354~g)   & 3034 $\pm~1.3$ &  3.6 $\pm~0.19$ &  9.6  $\pm~0.0001$ 
             & 14.5           \vspace{1mm} \\
Crystal 5 (390~g)   & 3034 $\pm~1.0$ &  5.0 $\pm~0.53$ &  8.5  $\pm~0.0003$ 
             & 16.9           \vspace{1mm} \\
Crystal 6 (340~g)   & 3034 $\pm~1.9$ &  3.5 $\pm~0.15$ &  6.5  $\pm~0.0002$ 
             & 12.1           \vspace{1mm} \\
\hline
\end{tabular*}
\end{table*}

The 50 events in the ROI that survived all of the selection requirements correspond to a background level of 0.55~counts/(keV$\cdot$kg$\cdot$y)~(ckky). These events were fitted with a linear background function plus a $0\nu\beta\beta$ signal term. Using the energy calibration and the response function for a mono-energetic $\gamma$ peak, the signal model function for each crystal is taken as the first term in Eq.~(\ref{equation_EMGExp}) with parameters of $\mu_{0\nu\beta\beta}$, $\sigma_{0\nu\beta\beta}$, and $\lambda_{0\nu\beta\beta}$. 
The uncertainty of $\mu_{0\nu\beta\beta}$ at the $Q$ value, denoted as $\epsilon_{\mu_{0\nu\beta\beta}}$, reflects the uncertainty in the energy calibration. The value of the skewness parameter $\lambda$ is scaled from the calibration data with the assumption that it is inversely proportional to the energy $\mu$. The standard deviation   $\sigma$ of the EMG function also depends on the energy.
The value of $\sigma_{0\nu\beta\beta}$ at the $Q$ value is determined from  the $\sigma$ values of other peaks using the relation
\begin{equation}
    \sigma^2(E) =\sigma_b^2 + c_1 E + c_2 E^2~,
\label{equation_resolution}
\end{equation}
where $\sigma_b$ is the base resolution independent of the energy input, and $c_2$ represents a linear relation of energy resolution with respect to the energy. The value of $c_1$ represents an additional term arising from the non-linearity of the energy calibration and the combined effect from various other sources of $E^2$ dependence.
This $E^2$ term is attributed to a position dependence of the signal-amplitude in the crystal, imperfect phonon collection, and temperature fluctuations. 
The determined values of $\epsilon_{\mu_{0\nu\beta\beta}}$, $\sigma_{0\nu\beta\beta}$, and $\tau_{0\nu\beta\beta}=1/\lambda_{0\nu\beta\beta}$ are listed in Table~\ref{table_Parameters}. The normalized detector response function, $f_1$, for a mono-energetic signal in the presence of a constant background is:

\begin{equation}
\begin{aligned}
	f_{1}&(E_{i,j}~|~\mu_i,\sigma_i,\lambda_i,n_{i,s},n_{i,b})= \\
		& \Big( \frac{n_{i,s}}{n_{i,s}+n_{i,b}}\Big) \cdot f_\textrm{EMG}(E_{i,j};\mu_i,\sigma_i,-\lambda_i) + \Big(\frac{n_{i,b}}{n_{i,s}+n_{i,b}}\Big) \cdot \frac{1}{W}~,
\end{aligned} 
\label{equation_EMGFlat}
\end{equation}

\noindent
where the $n$ values are the expected numbers of events, and $W$ is the 300~keV ROI interval. The indices $s$ and $b$ denote the signal and the background, respectively; and the indices $i$ and $j$ indicate the crystal number and the event number for each crystal, respectively.

\subsection{Sensitivity to \zerodbd{} decay of \Mo{100}}

\begin{figure} 
\begin{minipage}{\columnwidth}
\centering
\includegraphics[width=8.4cm]{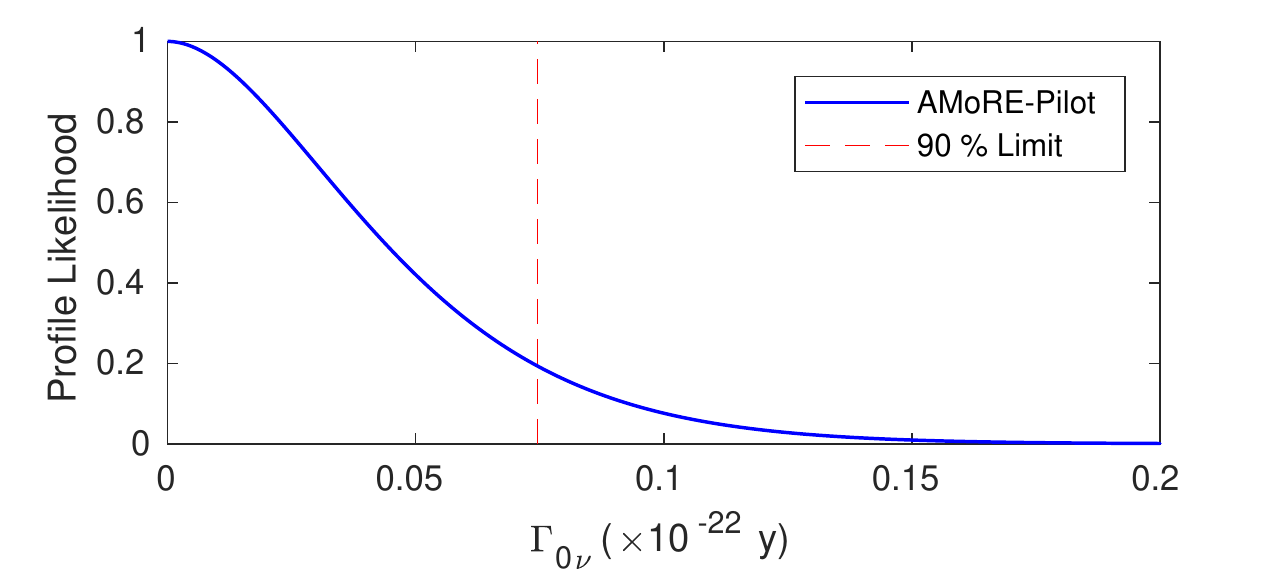}
\end{minipage}
\caption{Profile likelihood function with respect to the decay rate, $\Gamma_{0\nu}$. The dashed red line is the limit corresponding to 90\% confidence level. (Colour figure online)}
\label{figure_LikelihoodWErrors}
\end{figure}

Assuming Gaussian distributions for $\mu_i$, $\sigma_i$ and $\lambda_i$ described in Section 3, we marginalize the parameters in the likelihood function for each crystal. We then construct simultaneous unbinned extended likelihood functions for the six crystals and maximized the likelihood to fit the data. The best fit value for the decay rate $\Gamma_{0\nu}$ is zero and there is no evidence of an \Mo{100} \zerodbd{} signal in the current data. We evaluate 
 $\Gamma^{\textrm{Limit}}_{0\nu} < 7.3\times10^{-24} $~y$^{-1}$ at 90\% C.L. as indicated in Figure~\ref{figure_LikelihoodWErrors}. The corresponding half-life limit is: 
\begin{equation}
T^{0\nu}_{1/2} > 9.5\times10^{22}~\textrm{y}.
\end{equation}

\noindent
The solid green line in Figure~\ref{figure_ROISpectrum} is the combined response function for all six crystals with the parameters presented in Table~\ref{table_Parameters} and a \zerodbd{} decay rate corresponding to the 90\% limit of $9.5\times10^{22}~\textrm{y}$.

The 111~kg$\cdot$d (52.1~kg$\cdot$d of \Mo{100}) exposure and 0.55~ckky background level translates into an expected detection sensitivity for the present measurement of $T^{0\nu}_{1/2} > 1.1\times10^{23}$~y~\cite{feldman1998}, compatible with the experimental limit reported here.

The half-life limit reported here is near the projected experimental sensitivity of $1.1\times10^{23}$~y, and one order of magnitude lower than the current best limit for \Mo{100} \zerodbd{} decay: $1.1\times10^{24}$~y set by NEMO-3~\cite{arnold2015results}, and slightly better than the LUMINEU result of $7\times10^{22}$~y~\cite{lumineu2017limit} based on similar low-temperature thermal detectors and obtained with a 22~kg$\cdot$d of \Mo{100} exposure. Note that both the AMoRE-Pilot and the LUMINEU are in their development stages with small exposures. Both experiments are continuing to take follow-up measurements with lower backgrounds and larger detector masses, and are expected to make substantial improvements in their detection sensitivities.

\subsection{Neutrino mass limit}

The corresponding limit on the effective Majorana neutrino mass, $\langle m_{\beta \beta}\rangle$, is determined from Eq.~(\ref{equation_Thalf0nbb}). Using a phase space factor of 15.92$\times$10$^{-15}$ y$^{-1}$~\cite{Kotila2012PhaseSpace} and a representative value of $g_A = 1.27$, the corresponding  limit on the effective Majorana mass is $\langle m_{\beta\beta} \rangle~<~1.2$-$2.1$~eV. The uncertainty of the $\langle m_{\beta \beta} \rangle$ limit arises from the wide range of values of nuclear matrix element calculated with different techniques~\cite{simkovic2013QRPA,Vaquero2013NEDF,barea2015NME, Hyvarinen2015QRPA,yao2017REDF, rath2017NME}. 
The same calculation gives a detection sensitivity limit of 1.1-1.9~eV.

\section{Discussion and Conclusions}

In summary, we find no evidence of \zerodbd{} decay of \Mo{100} with a live exposure of 111~kg$\cdot$d using \enCMO{} crystals. An upper limit of $T^{0\nu}_{1/2} > 9.5\times10^{22}$~y is achieved with a background level of 0.55~counts/(keV$\cdot$kg$\cdot$y), which will be improved in succeeding measurements. 

The result reported here is the first low temperature measurement using MMC sensors with \enCMO{} crystals for \zerodbd{} search experiment. Although this is not the best limit on \zerodbd{} decay of \Mo{100}, it provides validation of the AMoRE detection concept for a competitive \zerodbd{} search. The background levels seen in the present study likely have external origins, and  major changes are being incorporated in the shielding arrangement to reduce them. Using the levels of radioactive contaminants measured in a series of HPGe counting measurements for the materials used in the detector assembly, simulations indicate the presence of non-negligible backgrounds that originate from the materials used for sensor readout, such as pin connectors, printed circuit boards~(PCBs), and Stycast-2850 epoxy. During the preparations for the sixth AMoRE-Pilot run, 
Polyimide-based PCBs were checked for their radioactive contamination levels and used to replace the G10 PCBs that were used in the fifth run. In addition, no pin-connectors or Stycast epoxy were used in cryostat regions below the mixing chamber plate (below the low-background shield layers of 10~cm thick lead and 6~cm thick copper). The MC simulations suggest that almost all of the nearly constant backgrounds between 4~MeV and 8~MeV in the electron spectrum are caused by neutron captures in the cryostat and the surrounding structures. To reduce the neutron-induced background, we installed layers of neutron shields in the detector setup. The neutron shields include a layer of 4~mm boric acid placed between the outer vacuum chamber and the lead shield, 2.5~cm thick borated-polyethylene plates, and 10-30~cm thick layers of polyethylene bricks surrounding the sides, bottom and top of the detector system. Background levels that remain after these modifications are being investigated.

The detector setup with MMC readout is designed to provide fast response time for heat and light detection. The light signals for $\alpha$ and $\beta/\gamma$ events have fast and slow components~\cite{gbkim_trans}. The time response of the MMCs to the fast component is close to the intrinsic rise-time of approximately 200~$\mu$s~\cite{hjlee2015}, which is one to three orders of magnitude faster than those for the thermal detectors based on NTD-Ge thermistor readouts~\cite{cuore0,pedretti2004single,Armengaud2017,Azzolini2018}. It is an especially important feature because of the relatively short \twodbd{} half-life of \Mo{100} (7.1$\times$10$^{18}$~y)~\cite{BARABASH201552}. As AMoRE is scaled up to larger detector masses, unresolved pileups of two random \twodbd{} signals will be an irreducible major background~\cite{beeman2012plb,chernyak_random_coin_dbd}. The rate of unresolved pileups from \Mo{100} decay will be approximately 0.4 events/y in a 0.4~kg \enCMO{} with a pile-up rejection time resolution of 1~ms, which would be a non-negligible background for a larger number of crystals with a poorer time resolution. Thus, the use of MMCs as the temperature sensors has a clear advantage: better time resolution results in reduced background levels from unresolved pileup signals in rare event searches.

After the completion of the Pilot runs, AMoRE-I will be implemented with 18~crystals. These include seven additional \enCMO{} that are already on hand. In addition, five \Mo{100}-enriched Li$_2$MoO$_4$ and Na$_2$MoO$_4$ crystals that are currently being fabricated. The total mass of the AMoRE-I crystals will be approximately 6~kg~(3~kg of \Mo{100}). The internal backgrounds and detector performance in phonon-scintillation detection will be investigated to determine the type of crystals to be used for the AMoRE-II, the main phase of the project with a total crystal mass of 200~kg. AMoRE-II aims at improving the effective Majorana neutrino mass sensitivity to 20-50 meV. It will be installed in Yemi lab, a new 1,000 m deep underground laboratory that is under construction in Korea and will be available in 2020.

\section{Acknowledgement}
This research is supported by Grant no. IBS-R016-D1 and IBS-R016-G1. One of author (VK) was also supported in the framework of the Moscow Engineering Physics Institute Academic Excellence Project (contract 02.a03.21.0005, August 27, 2013). The group from the Institute for Nuclear Research (Kyiv, Ukraine) was supported in part by the program of the National Academy of Sciences of Ukraine  "Fundamental research on high-energy physics and nuclear physics (international cooperation)."
\newpage

\end{document}